\documentclass[conference]{IEEEtran}
\pdfoutput=1
\IEEEoverridecommandlockouts
\usepackage{cite}
\usepackage{amsmath,amssymb,amsfonts}
\usepackage{breqn}
\usepackage{float}
\usepackage{algorithmic}
\usepackage{caption}
\usepackage{graphicx}
\usepackage{textcomp}
\usepackage{xcolor}
\usepackage{hyperref}
\def\BibTeX{{\rm B\kern-.05em{\sc i\kern-.025em b}\kern-.08em
    T\kern-.1667em\lower.7ex\hbox{E}\kern-.125emX}}
\renewcommand{\baselinestretch}{1.3}

\begin{document}

\title{Analysis and Control of a Planar Quadrotor}

\author{
\textbf{Praveen Venkatesh*, Sanket Vadhvana*, Varun Jain*} \\
Indian Institute of Technology, Gandhinagar \\
Discipline of Electrical Engineering \thanks{*All authors contributed equally}}

\maketitle

\begin{abstract}
In this paper, we model the planar motion of a quadcopter, and develop a linear model of the same. We perform stability analysis of the open loop system and develop a PD controller for its position control. We compare the closed loop response between the linear and non linear systems using the controller developed for the linear system. We also perform stability analysis for the linear and non linear systems, and compare the PD controller with modern deep neural methods based on reinforcement learning actor-critic networks.
\end{abstract}

\begin{IEEEkeywords}
Quadcopter, Transfer Function, PID Controller  
\end{IEEEkeywords}




    
    


\section{Introduction}
A quadcopter or quadrotor is a flying machine that uses four rotors to stabilize itself during flight. Quadcopters have been increasingly used in commercial and industrial applications in recent times. Due to its flexible, and inexpensive nature, it serves as an excellent platform to perform a variety of tasks.  One of the major challenges in the construction of the drones and autonomous vehicles is the design of the controller. However, due to its simple construction, developing a mathematical model and designing a controller for the quadcopter is quite easy.

In this paper, we model the planar motion of a quadcopter and design a controller for position control of the quadcopter. Due to the symmetric nature of the vehicle, the planar motion study in this paper can be easily extended for all three dimension to achieve full control of the quadcopter.\\

All of our code along with an animation of the quadcopter can be found in this GitHub repository : \href{https://www.github.com/praveenVnktsh/Modelling-Control-and-Simulation-of-a-Planar-Quadcopter}{\textit{Link}}

\section{Mathematical Modelling}

\begin{figure}[h!]
    \centering
    \includegraphics[width = \linewidth]{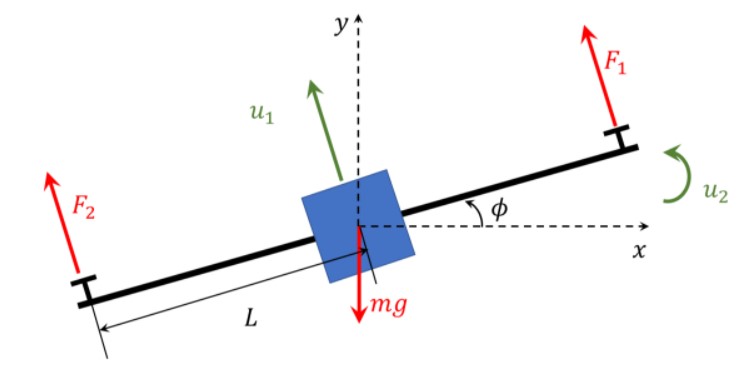}
    \caption{Quadcopter Model}
    \label{quad}
\end{figure}
\subsection{Qualitative Explanation}

The quadcopter is an inherently unstable system. Taking the simplified model seen in fig \ref{quad}, let us write the equations of motion.

Let $F_1$ and $F_2$ be the thrust forces exerted by the two motors on the frame. These are always perpendicular to the frame of the quadcopter. We hence define $u_1$ as the net thrust force exerted by the two motors, perpendicular to the frame of the quadcopter. Thus, we get the relation: 
\begin{equation} 
    u_1 = F_1 + F_2 
\end{equation}

Since the forces act at different locations on the same frame, there will be a net torque exerted by the two forces. We hence define $u_2$ as the net moment about the center of the frame as:

\begin{equation} 
    u_2 = \frac{L}{2}(F_1 - F_2 )
\end{equation}

Since we need to control the position of the quadcopter, the only inputs that can be controlled are the forces applied by the motors. We define $u = \begin{bmatrix} u_1 & u_2 \end{bmatrix}$ as the input vector of the system. As a result, the output vector of the system is $y_{out} = \begin{bmatrix} x & y \end{bmatrix}$. We do not consider the angle of the quadcopter as an output since at hover (steady state), the angle is always expected to be 0.

We do not model the thrust exerted by motors as coming from a spinning motor, but as simple forces being applied at the given location. We make this assumption since the quadcopter cannot move in and out of the plane, and the angular momentum offered by the spin of the motor does not affect other parameters of the system.

The other physical parameters of the system are listed in Table \ref{tab:my_label}.
\begin{table}[h!]
    \centering
    \begin{tabular}{||c|c|c||}
        \hline m & Quadrotor mass  & $0.18 Kg$ \\
        \hline g & Gravitational acceleration  & $9.8 m/s$ \\
        \hline L & Quadrotor span & 0.086 m \\
        \hline J & Quadrotor moment of Inertia & $2.5 \times 10^{-4} \mathrm{Kgm}^{2}$ \\
        \hline
    \end{tabular}
    
    \caption{Physical Parameters}
    \label{tab:my_label}
\end{table}

\subsection{Equations of Motion}
Having defined the input-output vectors, and the necessary parameters of the system, let us now model the system mathematically.

Writing Newton's 2nd Law of motion in the horizontal and vertical directions, we get:
\begin{equation}
    \Ddot{x} = -\frac{u_1sin({\phi})}{m}
\end{equation}
\begin{equation}
    \Ddot{y} = -g + \frac{u_1cos({\phi})}{m}
\end{equation}

\noindent Writing the torque equation about the center of mass, we get:
\begin{equation}
    \Ddot{\phi} = \frac{u_2}{J}
\end{equation}

In a compressed matrix form, the equations can be represented as:
\begin{equation}
    \begin{bmatrix}
    \Ddot{x}\\
    \Ddot{y}\\
    \Ddot{\phi}
    \end{bmatrix}
    =
    \begin{bmatrix}
    -\frac{sin{\phi}}{m} & 0 & 0\\
    \frac{cos{\phi}}{m} & 0 & -1\\
    0 &\frac{1}{J} & 0
    \end{bmatrix}
    \begin{bmatrix}
    u_1\\
    u_2\\
    g
    \end{bmatrix}
\end{equation}

\subsection{Determining the Equilibrium Point}

To determine the equilibrium point of the system, let us first define the state vector of the system
\begin{equation*}
    X = 
    \begin{bmatrix}
    x, y, \phi, \Dot{x}, \Dot{y}, \Dot{\phi}\end{bmatrix} ^ T 
\end{equation*}

where $x$, $y$ are the coordinates, $\phi$  is the angle, $\Dot{x}$, $\Dot{y}$ are the linear velocities, and $\Dot{\phi}$ is the angular velocity.

At equilibrium, since the quadcopter does not move, the state vector becomes:
\begin{equation*}
    X = 
    \begin{bmatrix}
   x, y, \phi, 0, 0, 0\end{bmatrix} ^ T 
\end{equation*}

\noindent The laws of motion hence become:
\begin{equation*}
    \begin{bmatrix}
    0\\
    0\\
    0
    \end{bmatrix}
    =
    \begin{bmatrix}
    -\frac{sin{\phi}}{m} & 0 & 0\\
    \frac{cos{\phi}}{m} & 0 & -1\\
    0 &\frac{1}{J} & 0
    \end{bmatrix}
    \begin{bmatrix}
    u_1\\
    u_2\\
    g
    \end{bmatrix}
\end{equation*}

\noindent Solving the above matrix equation, we get the following:
\begin{align*}
u_1sin{\phi} &= 0 \\
u_1cos{\phi} &= mg\\
u_2 &= 0
\end{align*}
$u_1 sin\phi = 0$ is valid for all $u_1$ only if $\phi = 0$. Hence, we obtain our equilibrium point as:
\begin{equation}
    X = 
    \begin{bmatrix}
    0, 0, 0, 0, 0, 0
    \end{bmatrix} ^ T 
\end{equation}

Since there are no constraints on the coordinates of the quadcopter, as given in the problem statement, let us take the coordinates to be $(0, 0)$ for ease of analysis.

Since the equilibrium angle of the quadcopter is $\phi = 0$, the equilibrium state also forces the constraint $u_1 = mg$.

\subsection{Linearization}
Let us consider the net thrust $u_1$, net moment $u_2$ and gravity $g$ as the inputs and the $x$ and $y$ coordinates of the quadrotor to be the outputs. We choose $g$ as an input as it appears as a constant term in the equations, which can otherwise not be represented in the state space equations. Now, let us linearize this equation:
\begin{align*}
    \Ddot{x} = -\frac{u_1sin({\phi})}{m} &= f_1(x,y,\phi,\dot{x},\dot{y},\dot{\phi},u_1,u_2,g) \\
    \Ddot{y} = -g + \frac{u_1cos({\phi})}{m} &= f_2(x,y,\phi,\dot{x},\dot{y},\dot{\phi},u_1,u_2,g) \\
    \Ddot{\phi} = \frac{u_2}{J} &= f_3(x,y,\phi,\dot{x},\dot{y},\dot{\phi},u_1,u_2,g)
\end{align*}

\begin{multline*}
\delta\ddot{x} =
\left.\frac{\partial f_{1}}{\partial x}\right|_{\text {EQ}}\delta x 
+ \left.\frac{\partial f_{1}}{\partial y}\right|_{\text {EQ}}\delta y 
+\left.\frac{\partial f_{1}}{\partial \phi}\right|_{\text {EQ}} \delta \phi \\
+ \left.\frac{\partial f_{1}}{\partial \dot x}\right|_{\text{EQ}} \delta\dot{x} 
+ \left.\frac{\partial f_{1}}{\partial \dot y}\right|_{\text{EQ}} \delta\dot{y} 
+ \left.\frac{\partial f_{1}}{\partial \dot\phi}\right|_{\text{EQ}} \delta\dot{\phi} 
\end{multline*}

\begin{equation}
    \Rightarrow \delta\ddot{x} = -9.8\times \delta\phi
\end{equation}

Similarly, we get the following equations after linearizing all of the Newtonian equations in a similar manner:
\begin{align}
    \delta\ddot{y} &= -\delta g + \frac{\delta u_2}{m}\\
    \delta\ddot{\phi} &= \frac{\delta u_2}{J}
\end{align}

\section{State Space Representation}
Now that we have a linearized representation of the system, let us try to represent the system in state space form.

We obtain the state space representation of the system as:

\subsubsection{State-Equation}

\begin{dmath}
    \begin{bmatrix}
    \delta\Dot{x}\\
    \delta\Dot{y}\\
    \delta\Dot{\phi}\\
    \delta\ddot{x}\\
    \delta\ddot{y}\\
    \delta\ddot{\phi}\\
    \end{bmatrix} = 
    \begin{bmatrix}
    0 & 0 & 0 & 1 & 0 & 0\\
    0 & 0 & 0 & 0 & 1 & 0\\
    0 & 0 & 0 & 0 & 0 & 1\\
    0 & 0 & -9.8 & 0 & 0 & 0\\
    0 & 0 & 0 & 0 & 0 & 0\\
    0 & 0 & 0 & 0 & 0 & 0\\
    \end{bmatrix}
    \begin{bmatrix}
    \delta x\\
    \delta y\\
    \delta\phi\\
    \delta\Dot{x}\\
    \delta\Dot{y}\\
    \delta\Dot{\phi}
    \end{bmatrix}
    +
    \begin{bmatrix}
    0 & 0 & 0\\
    0 & 0 & 0\\
    0 & 0 & 0\\
    0 & 0 & 0\\
    \frac{1}{m} & 0 & -1\\
    0 &\frac{1}{J} & 0
    \end{bmatrix}
    \begin{bmatrix}
    \delta u_1\\
    \delta u_2\\
    \delta g
    \end{bmatrix}
\end{dmath}

\subsubsection{Output Equation}
\begin{equation}
    \begin{bmatrix}
    \delta x\\
    \delta y\\
    \end{bmatrix} =
    \begin{bmatrix}
    1 & 0 & 0 & 0 & 0 & 0\\
    0 & 1 & 0 & 0 & 0 & 0\\
    \end{bmatrix}
    \begin{bmatrix}
    \delta x\\
    \delta y\\
    \delta\phi\\
    \delta\Dot{x}\\
    \delta\Dot{y}\\
    \delta\Dot{\phi}
    \end{bmatrix}
    +
    \begin{bmatrix}
    0 & 0 & 0\\
    0 & 0 & 0\\
    \end{bmatrix}
    \begin{bmatrix}
    \delta u_1\\
    \delta u_2\\
    \delta g
    \end{bmatrix}
\end{equation}

From the state-space equations we obtain the transfer function matrix as:
\begin{equation}
    H(s) = C(sI - A)^{-1}  B + D
\end{equation}
Hence our open loop transfer function matrix becomes:
\begin{equation}
\Rightarrow H(s) = 
\left(\begin{array}{ccc} 0 & -\frac{g}{J\,s^4} & 0\\ \frac{1}{m\,s^2} & 0 & -\frac{1}{s^2} \end{array}\right)
\label{eqn:tf}
\end{equation}

From the open-loop transfer function matrix, we see that the \textit{poles} are at the origin and there are no \textit{zeros}.

After substituting the physical parameters of the quadcopter, our open loop transfer function matrix becomes:
\begin{equation}
H(s) = 
\left(\begin{array}{ccc} 0 & -\frac{39200}{s^4} & 0\\ \frac{50}{9\,s^2} & 0 & -\frac{1}{s^2} \end{array}\right)
\end{equation}

Please note that the output vector $Y_{out}(s) = H(s) \dot U(s)$, where $Y_{out}(s) = [X(s) \ Y(s)]^T$ is the output vector and $U(s) = [U_1(s) \ U_2(s) \ g/s]^T$ is the input vector in Laplace domain.

\section{Open Loop Analysis}

From our transfer function matrix, we notice that all of the transfer functions have poles at the origin with no zeros. Hence, the system is unstable. We also plot the inputs for a sinusoidal input and verify the same.
\begin{figure}[h!]
    \centering
    \includegraphics[width = \linewidth]{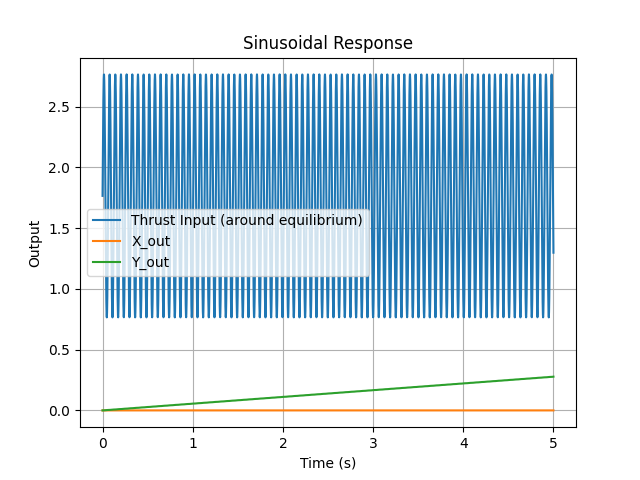}
    \caption{Sinusoidal Response - Thrust Input ($u_1$)}
    \label{fig:thrustopenloopu1}
\end{figure}

We can also see that the system parameters do not affect the location of the \textit{poles} and \textit{zeros}. Hence, even for quadcopters that have different construction, the design of the controller will remain same, with adjustments in the gain to account for differences in gains of the transfer function.

Now that we have established the mathematical model of the quadcopter, let us try to obtain its step response by solving the system of ODE's. 
\begin{figure}[h!]
    \centering
    \includegraphics[width = \linewidth]{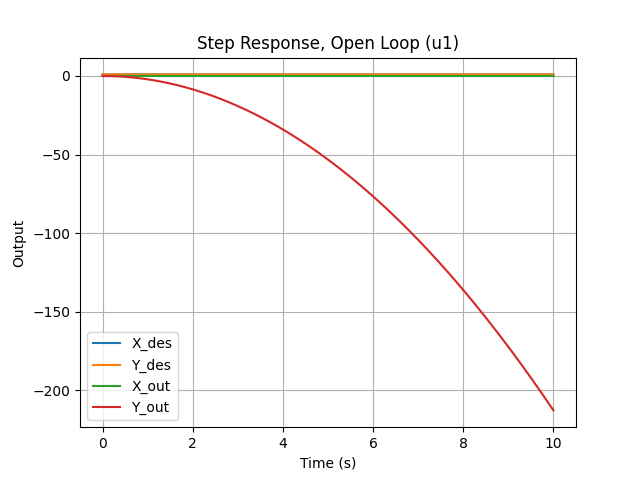}
    \caption{Open Loop Step Response - Thrust Input ($u_1$)}
    \label{fig:thrustopenloopu1}
\end{figure}

\begin{figure}[h!]
    \centering
    \includegraphics[width = \linewidth]{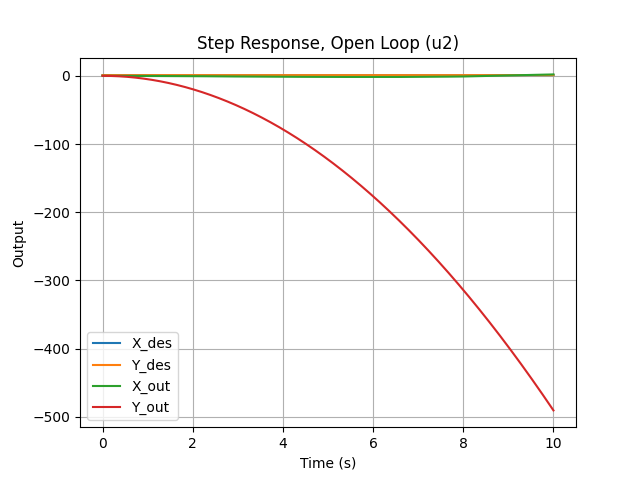}
    \caption{Open Loop Step Response - Tau Input ($u_2$)}
    \label{fig:thrustopenloopu2}
\end{figure}

\begin{figure}[h!]
    \centering
    \includegraphics[width = \linewidth]{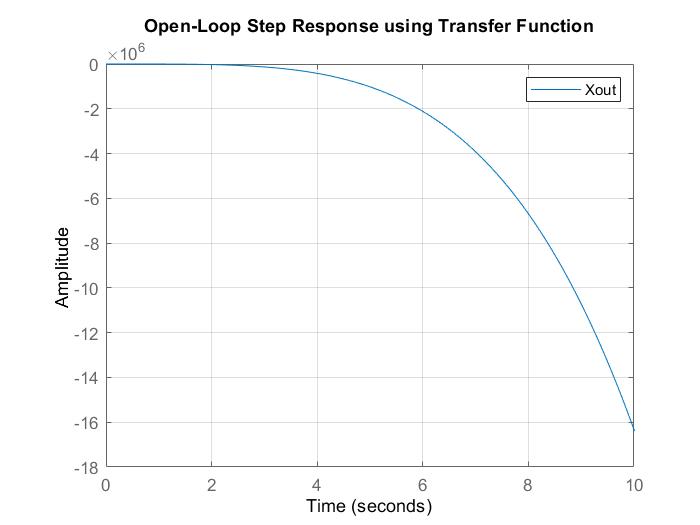}
    \caption{Open Loop Step Response using Transfer Function- Tau Input ($u_2$)}
    \label{fig:thrustopenlooptfu1}
\end{figure}

\begin{figure}[h!]
    \centering
    \includegraphics[width = \linewidth]{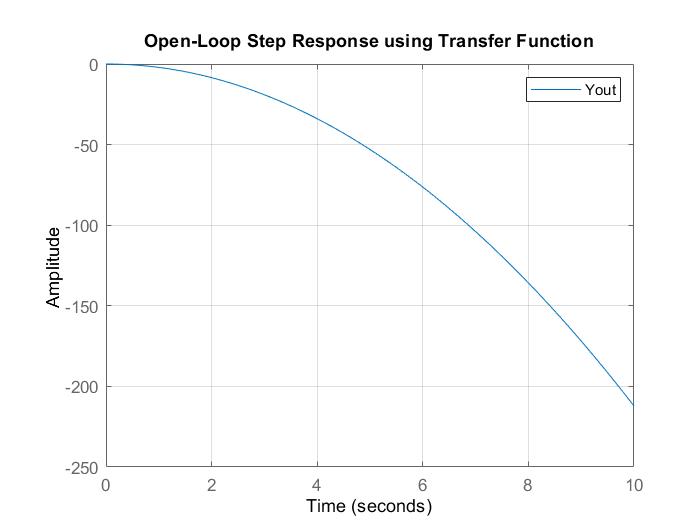}
    \caption{Open Loop Step Response using Transfer Function - Thrust Input ($u_1$)}
    \label{fig:thrustopenlooptfu2}
\end{figure}

From the open-loop step responses in figures \ref{fig:thrustopenloopu1} \& \ref{fig:thrustopenloopu2} we verify that the quadcopter is unstable, as previously predicted from the transfer functions. The output $(x, y)$ grows without bounds (in the negative direction), for a given bounded input (1, 0) or (0, 1), or (1, 1)

\begin{figure}
    \centering
    \includegraphics[width = \linewidth]{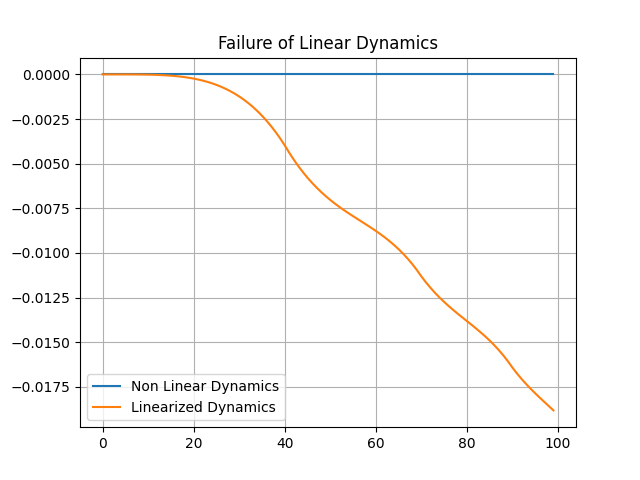}
    \caption{Linear Model Failure - x coordinate}
    \label{fig:linearfailx}
\end{figure}

\begin{figure}
    \centering
    \includegraphics[width = \linewidth]{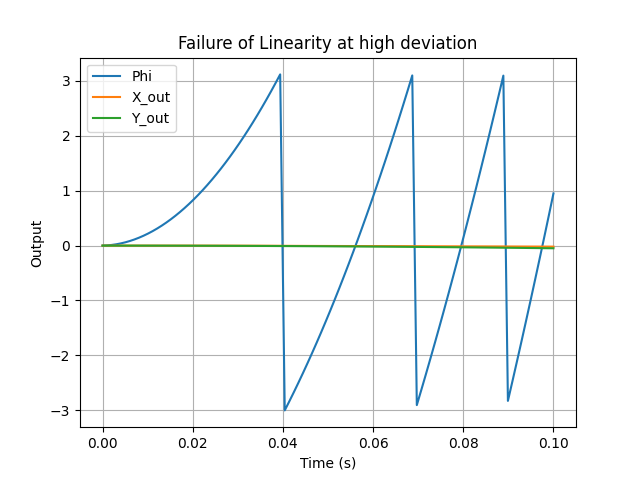}
    \caption{Linear Model Failure}
    \label{fig:linearfail}
\end{figure}

We also obtain the step responses from the transfer function matrix computed in equation \ref{eqn:tf}in figures \ref{fig:thrustopenlooptfu1} and \ref{fig:thrustopenlooptfu2}.
We notice that the step response is exactly the same as obtained from the ODE solver for the y-axis. However, there is a significant deviation between the graphs for the x-axis plots. This is because the linear approximation starts to fail very quickly as the angle of the quadcopter depends on the input $u_2$. Since $u_2$ is set to $1 N/m$ when evaluating the step response, the high values of angle obtained from the step response leads to a large deviation from linearity, causing the transfer function to fail. Hence there is a large deviation from what is observed in the ODE solver, and what the transfer function is predicting. This has been shown in the graph \ref{fig:linearfail} where we see that the angle rapidly keeps increasing and reaches $\pi \text{ rad}$ at 0.04 seconds, and the linear approximation starts failing. A comparative plot of the ODE solution for the x-coordinate for both linear and non linear systems is also seen in figure \ref{fig:linearfailx}.

\section{Controller Design}

Since the quadcopter is inherently an unstable system, we design a controller to stabilize the quadcopter.
Given a desired $(x_{des}, y_{des})$ coordinate pair, we want to design a controller that operates on the input vector $u = [u_1, u_2]$, and moves the quadcopter from $(x_{i}, y_{i})$ to $(x_{des}, y_{des})$

\begin{figure}
    \centering
    \includegraphics[width = \linewidth]{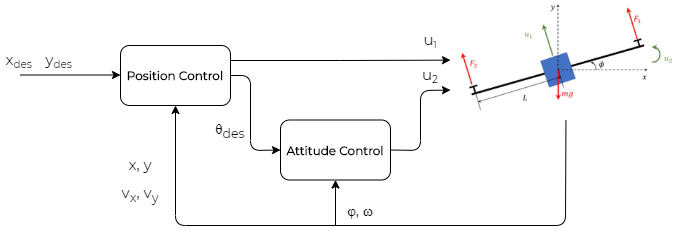}
    \caption{Control Mechanism}
    \label{fig:controllerblockdiag}
\end{figure}

A block diagram of the controller that we design can be seen in figure \ref{fig:controllerblockdiag}. We use PD controllers at each stage to control the coordinates and the angle of the quadcopter. It is evident from the transfer functions are type 4 (x-axis) and type 2 (y-axis). Hence, for any given compensator design with a maximum of 1 pole and 1 zero, the type of the system will never decrease. Hence, since the controllers can track velocity with a constant error (y-axis, type 2), and jerk (x-axis, type 4), there will be no steady state error. Hence, we can utilize PD controllers for the control of the quadcopter. 

Let us now try to intuitively understand the reason for the design of the two loops for the control. At an arbitrary angle $\phi$ of the quadcopter, there are two components of the thrust force $u_1$, one along the $y$ direction, and another along the $x$ direction. The components of force along these directions is dictated by the angle of the quadcopter, which is primarily controlled by the input moment $u_2$. Let us now assume that the quadcopter is initially at the origin at $(0, 0)$, and we want to take it to $(1, 1)$ (step response). It is intuitively clear that to increase the altitude upto a height of $1$ unit, the thrust should initially be high, and then rapidly decrease after it reaches the desired state. However, this should happen at the same time when the drone is moving to a horizontal distance of $1$ unit. Hence, we introduce another variable called $\phi _{control}$, which controls the desired angle required at each point of the drone. Intuitively, we can see that the desired angle of the drone needs to be $0$ at the destination point, and in the beginning, it needs to point towards the destination. Hence, we express the desired angle as a function of the error in the required $x$ coordinate, and the current $x$ coordinate. Therefore, we have a total of 3 controllers, one that controls the thrust (depending on the vertical error), one that controls the desired angle (depending on the horizontal error), and one that controls the moment (depending on the angle error).

We can also understand this choice by looking at the equations. We see that the angle of the quadcopter is solely determined by the input moment that is supplied, and the location of the quadcopter depends on the thrust force and the current angle of the quadcopter. Hence, we can see that a total of 3 controllers are required, one for each of the coordinates, and one for the desired angle at each point in space, depending on the coordinates.

Implementing the controllers, our final control equations become:

\begin{align}
    u_1 &= mg + K_{p1} (y_{des} - y) + K_{d1}(0 - \dot{y}) \\
    \phi _{des} &= K_{p2} (x_{des} - x) + K_{d2}(0 - \dot{x}) \\
    u_2 &=  K_{p2} (\phi _{des} - \phi) + K_{d2}(\dot{\phi_{des}} - \dot{phi})
\end{align}




\section{PD Tuning}

Since we use the ODE solver in order to execute our controller, we tune the gains depending on the oscillations that we observe in the system.
We follow the following procedure to tune the controller:
\begin{enumerate}
    \item Initially, we increase the $K_p$ parameter from 0 to some non zero value until we get sustained oscillations.
    \item Slowly increase the $K_d$ parameter to increase the damping of the system. We increase it until the system overshoot is within the limit.
    \item Increase the $K_p$ parameter to decrease the rise time to the desired value.
    \item Repeat for all axes.
\end{enumerate}

Using this method, we obtain the gains as follows:
\begin{enumerate}
    \item $K_{p1} = 7.0$
    \item $K_{d1} =  1.42$
    \item $K_{p2} = 2.5$
    \item $K_{d2} = 0.56$
     \item $K_{p2} = 0.04$
    \item $K_{d2} = 0.008$
\end{enumerate}

\section{Closed loop Stability}

\begin{figure}
    \centering
    \includegraphics[width = \linewidth]{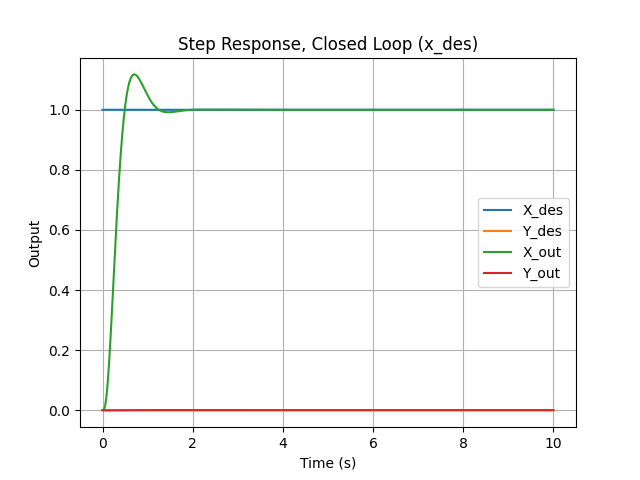}
    \caption{Closed Loop Step Response Xdes = 1 - Overshoot = 11.78\% \& Rise Time = 31.5ms}
    \label{fig:closedloopx}
\end{figure}
\begin{figure}
    \centering
    \includegraphics[width = \linewidth]{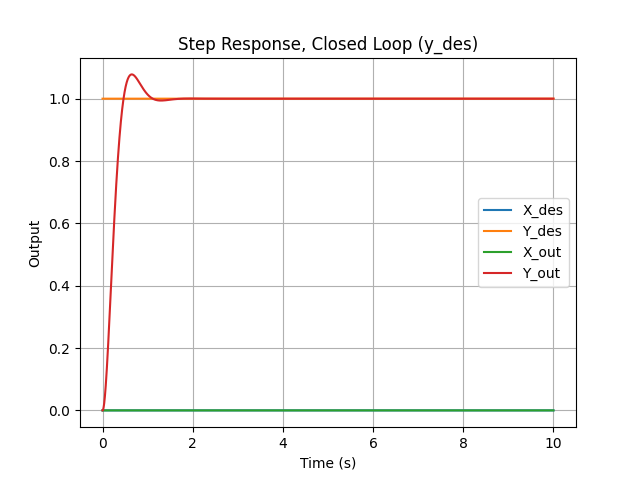}
    \caption{Closed Loop Step Response Ydes = 1 - Overshoot = 7.78\% \& Rise Time = 32.1ms}
    \label{fig:closedloopy}
\end{figure}

From the closed-loop step responses seen in figures \ref{fig:closedloopx} and \ref{fig:closedloopy}, we can see that the response is stable as for a bounded input, we get a bounded output. 

Let us now compare these plots with those of the open-loop step responses that we obtained previously. It is clear that the design objectives have been met. In the open loop response, the outputs grew without bounds. However, here we see that the output rises to the desired input and stabilizes quickly within the required overshoots and rise times, making the entire system stable.

\begin{table}[]
\centering
\begin{tabular}{|l|l|l|}
\hline
\textbf{Axis} & \textbf{Overshoot} & \textbf{Rise Time} \\ \hline
X             & 11.78\%            & 31.5ms             \\ \hline
Y             & 7.78\%             & 32ms               \\ \hline
\end{tabular}
\caption{System Results - Linear System}
\label{syspar}
\end{table}

From table,\ref{syspar} we can see that the specifications of the system has been met as the overshoot and rise time for both axes are within the limits of 16\% overshoot and $\omega _n > 5 rad/s$

\section{Non-Linear System}

We now use the PD controller developed for the linearized system, we try to utilize the controller on the non-linear model.

\subsection{Closed Loop Step Responses}

\begin{figure}
    \centering
    \includegraphics[width = \linewidth]{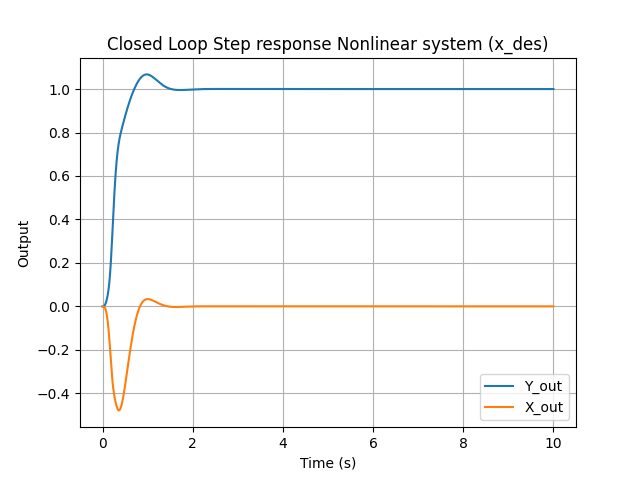}
    \caption{Non linear Closed Loop Step Response - X}
    \label{fig:nonlinstepx}
\end{figure}

\begin{figure}
    \centering
    \includegraphics[width = \linewidth]{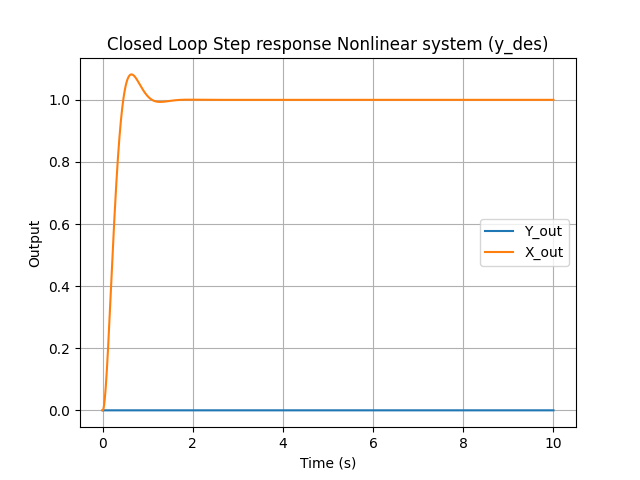}
    \caption{Non linear Closed Loop Step Response - Y}
    \label{fig:nonlinstepy}
\end{figure}

\begin{figure}
    \centering
    \includegraphics[width = \linewidth]{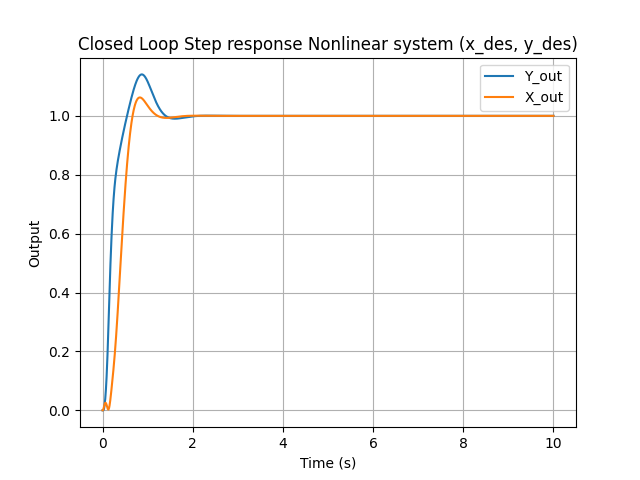}
    \caption{Non linear Closed Loop Step Response - X \& Y}
    \label{fig:nonlinstepxy}
\end{figure}

We notice that the step responses (figures \ref{fig:nonlinstepx} \ref{fig:nonlinstepy} \ref{fig:nonlinstepxy}) are quite similar to the ones obtained in the linear system. However, we notice that there are some slight deviations in the obtained graph compared to the linearized system. This is to be expected since the linear approximation of $sin\phi = \phi$ and $cos\phi = 1$ becomes invalid at high values of $\phi$.

\begin{table}[h!]
\centering
\begin{tabular}{|l|l|l|}
\hline
\textbf{Axis} & \textbf{Overshoot} & \textbf{Rise Time} \\ \hline
X             & 20.83\%            & 28ms               \\ \hline
Y             & 3.16\%             & 40ms               \\ \hline
\end{tabular}
\caption{System Results - Non linear system}
\label{tab:nonline}
\end{table}

We obtain the rise times and overshoots from table \ref{tab:nonline}. We see that the design objectives are not met. The overshoot for x axis increases and the rise time decreases, whereas the overshoot for y axis decreases and rise time increases. This is because $sin\phi < \phi$ for all values of $\phi$, which causes a repeated overestimation of the angle in the linear system leading to lower compensation in the x-axis, whereas for the same over-estimation, $cos\phi$ is closer to $\phi$ leading to higher compensation and hence an decrease in overshoot and increase in rise time for the y-axis.

\subsection{Stability Analysis}

To ascertain the stability of the system, let us try different initial states and try to see if the system achieves equilibrium.

\begin{figure}
    \centering
    \includegraphics[width = \linewidth]{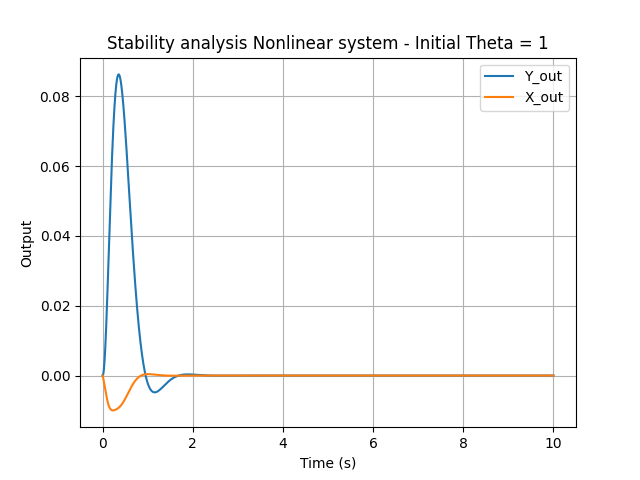}
    \caption{Non linear Closed Loop Step Response - $\phi _i = -0.5$}
    \label{fig:nonlinstab1}
\end{figure}

\begin{figure}
    \centering
    \includegraphics[width = \linewidth]{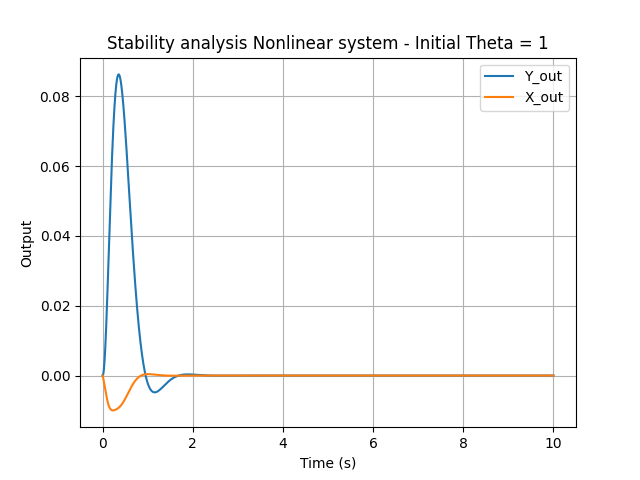}
    \caption{Non linear Closed Loop Step Response - $\phi _i = 1$}
    \label{fig:nonlinstab2}
\end{figure}

From the graphs (\ref{fig:nonlinstab1}, \ref{fig:nonlinstab2}) , it is quite clear that the system quickly comes back to equilibrium and stays there. It is hence quite clear that the system is stable.

\section{Simulation}

In order to simulate the system, we utilize python and the ODE solver available in scipy's library. Rendering is carried out using OpenCV. We choose python for our codebase as we also implement reinforcement learning algorithms for attitude control.

\begin{figure}[h]
    \centering
    \includegraphics[width = \linewidth]{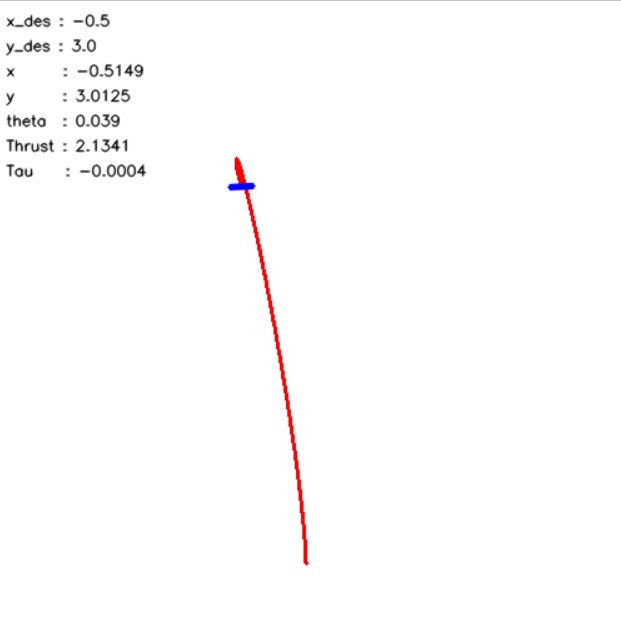}
    \caption{Simulator}
    \label{sim}
\end{figure}

A visual representation of the simulation is seen in fig. \ref{sim}.

\section{A Smart (not so smart) Controller}

\subsection{Reinforcement Learning}
Reinforcement Learning is a growing area of interest in Machine learning, which involves software agents(Controllers). These agents attempt to interact with the system in the hope of maximising a prioritized reward based on the feedback received by from the system. 

\begin{figure}[h]
    \centering
    \includegraphics[width = \linewidth]{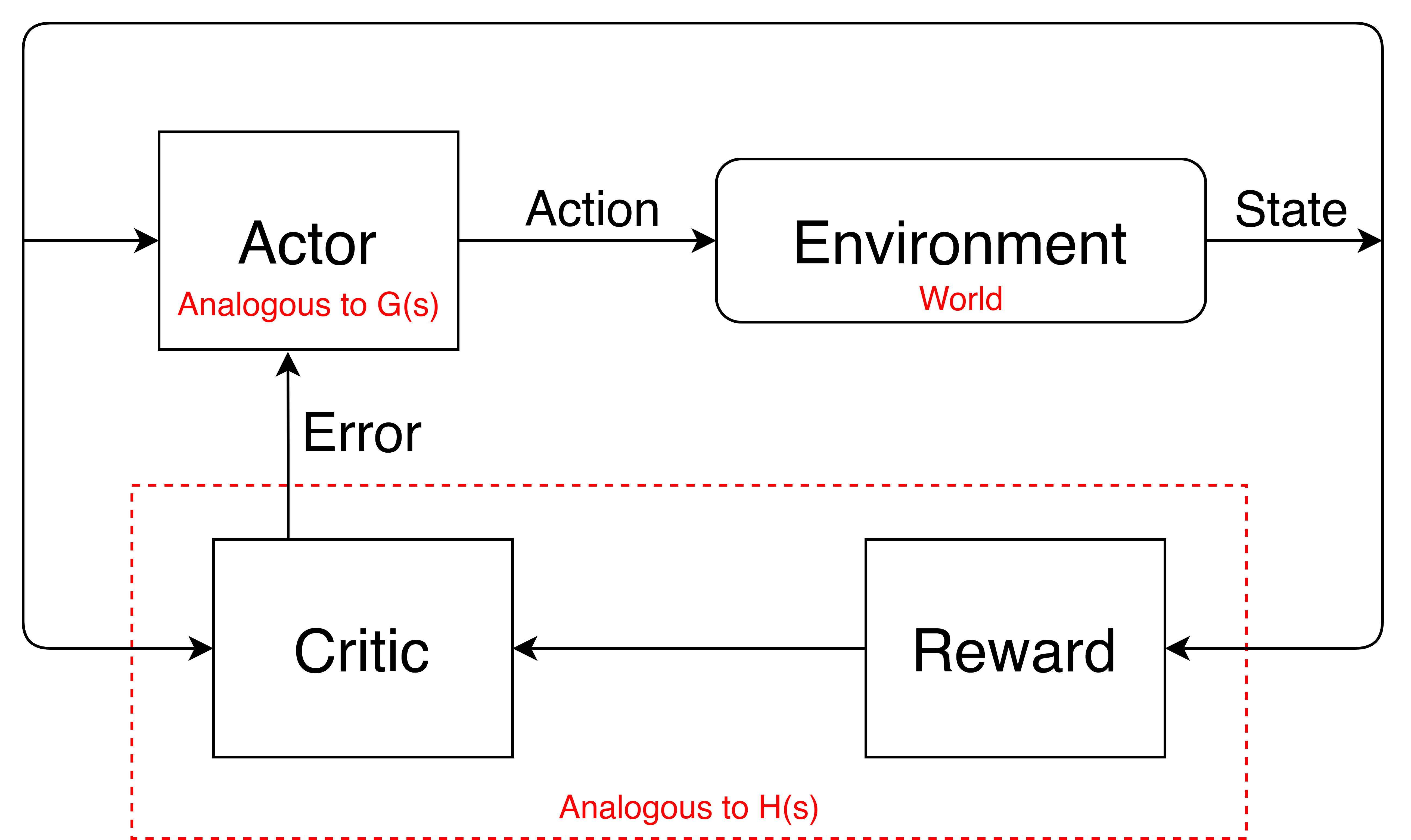}
    \caption{Reinforcement Learning}
    \label{rl}
\end{figure}

We can look at reinforcement learning methods as being analogous to the closed loop controller of a system. In reinforcement learning instead of having a constant transfer function G(s) and feedback H(s), we have a set of Policy functions namely known as the Actor and a Reward System called Critic as shown in the figure \ref{rl}. 

\subsection{Proximal Policy Optimization}

Proximal Policy Optimization\cite{DBLP:journals/corr/SchulmanWDRK17,ppo_openai} is a form of Policy Gradient method that rely on a Stochastic Policy which samples actions given the state of the agent in the environment based on a learnt probability distribution. 

The parameters for the algorithm are updated based on this update equation:
\begin{center}
    $\nabla_{\phi}E_{x}[f(x)] = E_{x}[f(x)\nabla_{\phi}logp(x)]$
\end{center}
where, $x$ is the possible action, $f(x)$ is the reward function and $p(x)$ is the probability of x in the probability distribution.

\subsubsection{Implementation}
We create an Environment/System for the agent to interact with using the OpenAI Gym API. Using the ODE solver this Environment simulates a Quadcopter. The environment gives our Agent(Controller) the ability to act with it using the API. The controller has no prior knowledge of the system that it is supposed to control. Based on random actions taken initially, it develops a sense of what it is supposed to do. Using reward shaping, we try to mould our agent to perform the task of controlling the drone towards a desired point.

Rewards that we are using are,
\begin{itemize}
    \item A distance function, that rewards it based on how far it is from the goal location.
    \item Negative reward for undesired behaviour such rotating about the axis without moving towards the goal
    \item Reward for maintianing the steady state position.
\end{itemize}

\begin{figure}[h]
    \centering
    \includegraphics[width = \linewidth]{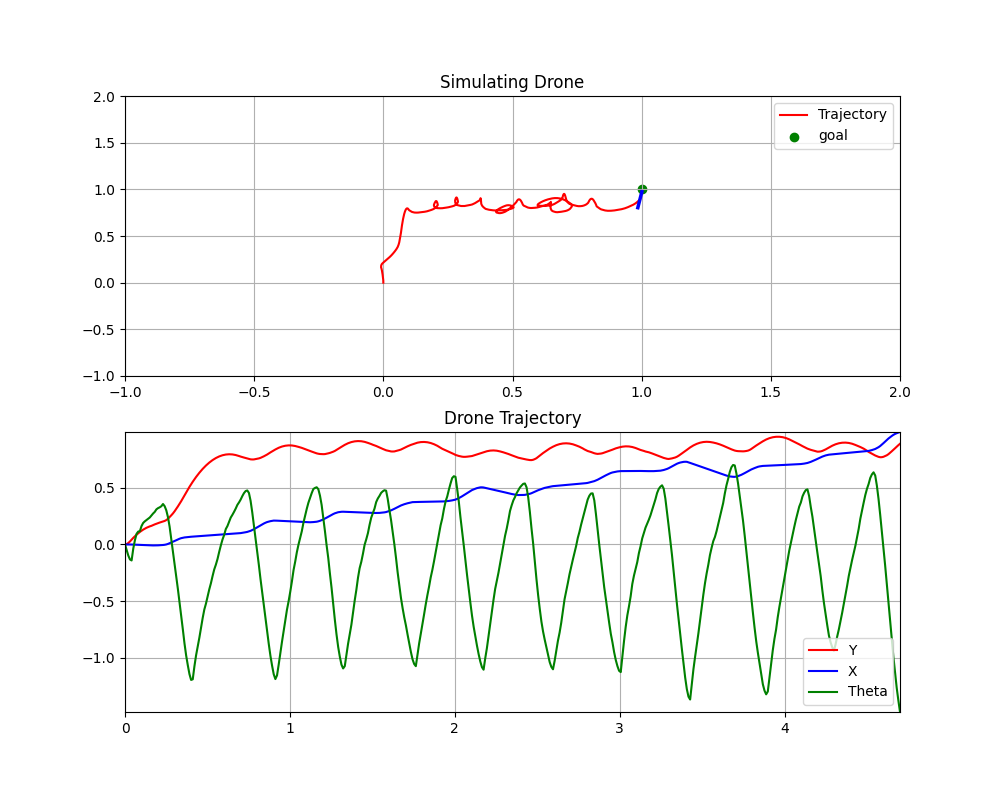}
    \caption{Closed loop step response for RL Model}
    \label{rlother}
\end{figure}

We observe that even though the PPO model is able to finally able to understand its goal of reaching the setpoint, the PD performs significantly better than the PPO model. Here we see how classical models sometimes beat more sophisticated systems. We plot closed loop step response and observe that the overshoot is much higher than that found after the PD tuning, and the trajectory is not smooth. (Fig \ref{rlother})

\section{Limitations, Recommendations and Outlook}

In this paper, we present the full dynamics of a planar quadrotor, and linearize it to develop a controller for the same. We design the controller for the described design criteria, and attempt to use it for the non linear system. We show that the unstable linear and non linear systems can both be stabilized using a 3 PD controllers with different gains.

As seen from the trajectories of the non-linear and linear systems, at angles that are deviating far from equilibrium, the controller is unable to adequately compensate the system to follow the ideal path. This is due to the linear conditions failing to be satisfied. Moreover, we have not modelled any environmental effects that may exist in a real environment such as wind and prop wash (when the drone flies close to the ground). In a real setting, to combat these effects, a PID controller with a very small $K_i$ is used to offset any errors caused due to environmental effects. We have also not modelled the motors themselves, but are rather considering the forces to be applied at the ends of the quadrotor. In a real quadrotor, the spin of the motors will increase the angular momentum of the system, and if left unbalanced will cause the drone to yaw unnecessarily.

The linear system analysis is valid only as long as the angle of the quadcopter is low. Hence, even though the PD controller may still work for the actual non linear system, designing the controller on the basis of the mathematical model may not guarantee that the design objectives are met. Hence, when designing such controllers, we need to add a safety margin to our specifications so that the non linear system can behave according to the required specifications.

In recent times, there has been an increasing interest in the design of controllers using deep learning techniques. In this paper, we also present a deep learning controller based on the PPO algorithm and show that deep learning algorithms can also be used to control the drone. However, these require a large amount of training data, and without such data, are inferior in performance compared to classical controllers such as PID.

From the graph, we also see some unusual behavior where the drone tries to rotate 360 degrees and control thrust to move towards the required point. This cannot be achieved in real life, and hence is an undeployable in a real drone!


\nocite{*}
\bibliographystyle{ieeetr}
\bibliography{main}

\end{document}